\documentclass[english,aps, prl, twocolumn, reprint, superscriptaddress]{revtex4}
\usepackage[T1]{fontenc}
\usepackage[latin9]{inputenc}
\usepackage{units}
\usepackage{amsmath}
\usepackage{graphicx}

\makeatletter

\@ifundefined{textcolor}{}
{%
 \definecolor{BLACK}{gray}{0}
 \definecolor{WHITE}{gray}{1}
 \definecolor{RED}{rgb}{1,0,0}
 \definecolor{GREEN}{rgb}{0,1,0}
 \definecolor{BLUE}{rgb}{0,0,1}
 \definecolor{CYAN}{cmyk}{1,0,0,0}
 \definecolor{MAGENTA}{cmyk}{0,1,0,0}
 \definecolor{YELLOW}{cmyk}{0,0,1,0}
}

\makeatother

\usepackage{babel}
\begin{document}

\title{Spectral interferometric polarised coherent anti-Stokes Raman spectroscopy}

\author{Brad Littleton}

\email{bradley.littleton@kcl.ac.uk}

\selectlanguage{english}%

\affiliation{Department of Physics, King's College London, Strand, London, WC2R
2LS}

\author{Thomas Kavanagh}

\affiliation{Department of Physics, King's College London, Strand, London, WC2R
2LS}

\author{Frederic Festy}

\affiliation{Biomaterials, Biomimetics and Biophotonics Department, King\textquoteright{}s
College London Dental Institute, Floor 17 Tower Wing, Guy\textquoteright{}s
Hospital, London Bridge, London SE1 9RT, UK}

\author{David Richards}

\email{david.r.richards@kcl.ac.uk}

\selectlanguage{english}%

\affiliation{Department of Physics, King's College London, Strand, London, WC2R
2LS}

\begin{abstract}
We have developed an interferometric implementation of coherent anti-Stokes
Raman scattering (CARS) which enables broadband coherent Raman spectroscopy
free from non-resonant background (NRB), with a signal strength proportional
to concentration. Spectra encode mode symmetry information into the
amplitude response which can be directly compared to polarised spontaneous
Raman spectra. The method requires only passive polarisation optics
and is suitable for a wide range of laser linewidths and pulse durations.
\end{abstract}
\maketitle
Spontaneous Raman scattering provides a powerful optical route to
obtain chemically specific information. Its application to microscopy
enables imaging using molecular vibrations as a contrast mechanism
and mapping of cell and tissue constituents based on their chemical
signature \citep{Stewart2012,Puppels1990}. Coherent Raman scattering
(CRS) is the non-linear multiphoton equivalent and allows much faster
acquisition, \citep{Saar2010,Parekh2010} with intrinsic optical sectioning
in imaging \citep{Cheng2004}. However, while CRS imaging has been
very successfully employed for imaging individual Raman bands, it
struggles to harness the powerful chemical specificity provided by
spectral detection in the Raman fingerprint region. In particular,
quantitative broadband coherent Raman spectroscopy of biological samples,
analogous to spontaneous Raman spectroscopy, has proven difficult
to achieve primarily due to the coherent backgrounds inherent to CRS.
In this letter we report on a new method for quantitative broadband
CRS spectral imaging, which uses passive polarisation optics combined
with spectrally-resolved balanced homodyne detection. The technique
has relaxed requirements on spectral phase and instrument stability
and provides full access to the Raman fingerprint region, while retaining
the advantages of enhanced signal and optical sectioning inherent
to CRS.

\begin{figure}
\begin{centering}
\includegraphics{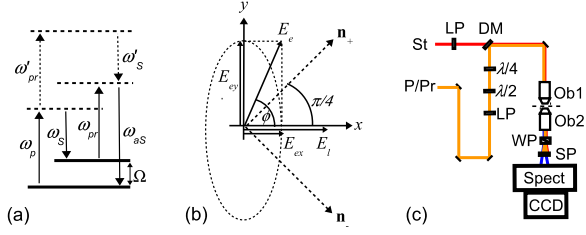}
\par\end{centering}

\caption{(colour online) (a) Energy level diagram. CARS involves the path $\omega_{p}\rightarrow\omega_{S}\rightarrow\omega_{pr}\rightarrow\omega_{aS}$,
SRS is $\omega_{p}\rightarrow\omega_{S}\rightarrow\Omega$ (where
$\Omega$ is a vibrational resonace). An example non-resonant FWM
pathway is shown as $\omega_{p}\rightarrow\omega'_{pr}\rightarrow\omega'_{S}\rightarrow\omega_{aS}$.
(b) Polarisations for SIP-CARS with one elliptical (or circular) polarised
field $E_{e}$ and one linear field $E_{l}$. Detection is performed
along $\mathbf{n}_{+}$ and $\mathbf{n}{}_{-}$. $E_{e}$ is shown
as the input to a $\nicefrac{\lambda}{4}$ plate (orientated at angle
$\phi$ to the fast axis) to generate the desired elliptical polarisation.
For a circular Stokes/linear pump: $E_{e}=E_{S},\, E_{l}=E_{p}=E_{pr},\,\phi=\nicefrac{\pi}{4}$.
For elliptical pump/linear Stokes: $E_{e}=E_{p}=E_{pr},\, E_{l}=E_{S},\,\phi=\nicefrac{\pi}{8}\,\mathbf{\mathrm{or\,}}\nicefrac{3\pi}{8}.$
(c) Experimental setup. Both pump/probe (P/Pr: 1050 nm) and Stokes
(St: 1070-1600 nm) beams were derived from a commercial supercontinuum
source with two outputs (repetition rate 20MHz, $\sim$5ps pulses
at the sample). Beams were combined on a dichroic mirror (DM) and
focussed into the sample by a 1.3NA oil immersion DIC objective lens
(Ob1), where the power of each beam was 30 mW. Samples were prepared
between two coverslips, separated by <60 $\mathrm{\mu m}$. The signal
was collected by a 0.75NA air objective (Ob2), separated from the
excitation fields by a shortpass filter (SP), and then dispersed via
grating spectrometer onto a deep depletion CCD. The pump/probe ellipticity
was controlled by a combination of linear polariser (LP) and zero-order
half- and quarter-waveplates ($\nicefrac{\lambda}{2}$ and $\nicefrac{\lambda}{4}$,
respectively). A Wollaston prism (WP) was placed after Ob2 to separate
the two detection polarisations on the CCD. The beams were again circularly
polarised before the spectrometer by an achromatic quarter-wave plate
(not shown) to account for the the different grating reflectivity
for each polarisation.\label{fig:1}}
\end{figure}

The two most widely used CRS techniques are stimulated Raman scattering
(SRS) and CARS (Fig \ref{fig:1}(a)) \citep{Eesley1981}. At intensities
suitable for biological samples, SRS requires heterodyne methods \citep{Freudiger2008,Slipchenko2012}
to detect the signal against the coherent background of the excitation
fields; such methods require wavelength scanning to build up spectral
information. CARS is a four wave mixing (FWM) process with a signal
field generated at the anti-Stokes frequency $\omega_{aS}=\omega_{pr}+\omega_{p}-\omega_{S}$
which is spectrally separated from the excitation fields ($\omega_{p},\,\omega_{S},\,\omega_{pr}$
are denoted the pump, Stokes and probe fields, respectively). CARS
can therefore simultaneously generate an entire vibrational spectrum
by using a spectrally broad Stokes beam \citep{Kano2005}, making
CARS suitable for rapid detection of vibrational spectra, particularly
in microscopy. 

The coherent background in CARS arises from FWM processes that are
independent of vibrational transitions (Fig \ref{fig:1}(a)). The
anti-Stokes intensity is determined by the third order susceptibility,
which (away from electronic resonances) is given by
\begin{equation}
\left|\chi\right|^{2}=\chi_{_{NR}}^{2}+2\chi_{_{NR}}\textrm{Re}\left\{ \chi_{_{R}}\right\} +\left|\chi_{_{R}}\right|^{2},\label{eq:1}
\end{equation}
where $\chi_{R}$ and $\chi_{NR}$ are components resonant and non-resonant
with vibrational modes, respectively. Since $\chi_{R}$ is proportional
to the number of resonant modes in a medium, the concentration dependence
of CARS is quadratic and non-linearly mixed with the non-resonant
background (NRB). The interferometric (second) term, however, is linear
in $\chi_{R}$ and is amplified by $\chi_{NR}$. Re\{$\chi_{R}\}$
is dispersive and antisymmetric about the vibrational line-centre;
however, $\mathrm{Im}\{\chi_{_{R}}\}$ is directly related to the
spontaneous Raman spectrum \citep{Eesley1981}.

Many methods have been developed to remove the effect of the NRB from
CARS spectra \citep{Day2011}. Non-interferometric techniques, such
as polarisation-based \citep{AKHMANOV1977,Oudar1979,Lu2008} and time-resolved
methods \citep{Volkmer2002,Lee2008}, recover a (typically small)
proportion of the energy in the third term of (\ref{eq:1}). However,
the interferometric term is the largest source of signal for weak
resonances and, below the damage threshold of biological samples,
CRS has been shown to be faster than spontaneous Raman spectroscopy
only if this term is detected \citep{Parekh2010,Cui2009}.

Im\{$\chi_{R}$\} may be recovered through interference between the
anti-Stokes field and a local oscillator (LO), which may be generated
either externally\citep{Potma2006,Muller2009,Jurna2008,Orsel2010a},
or by using the NRB as an internal LO\citep{Kee2006,Lim2005,Oron2003}.
Broadband external approaches acquire an entire spectrum simultaneously
via spectral interferometry, requiring high stability\citep{Evans2004}
and broadband transform limited laser sources\citep{Chowdary2010}
(precluding the use of fields generated in photonic crystal fibres
(PCF), which are typically much broader). Using the NRB as the LO
ensures that the phase relationship between the LO and anti-Stokes
fields is fixed, which relaxes stability and spectral coherence requirements;
however, as the LO is in phase with $\mathrm{Re}\{\chi\}$, recovering
$\mathrm{Im}\{\chi\}$ is not trivial. Previous experimental implementations
also need broad transform limited pulses \citep{Lim2005,Sung2011,Oron2003,Oron2002a,Wipfler2012},
or require phase scanning \citep{Kee2006,Wipfler2012}.

Computational internal LO approaches have also been developed, which
use the $\chi_{_{NR}}$ Re\{$\chi_{_{R}}$\} term present in raw CARS
spectra to calculate Im\{$\chi_{_{R}}$\} \citep{Vartiainen2006,Liu2009a}.
These produce good approximations to Im\{$\chi_{R}$\} if the spectra
are of sufficient width, however both methods intrinsically produce
a spectrally varying error signal, which increases at resonances (up
to 10\%)\citep{Cicerone2012}. Both techniques also require prior
or post estimation of the spectral variation of the NRB (effectively
the spectral variation of the Stokes field) from a reference material,
which is often compromised by the presence of a residual resonant
response. Changes of the background spectrum during an acquisition
leads to errors that can mask the weaker resonances in the fingerprint
region of biological samples\citep{Parekh2010}. 

We describe here a new internal LO technique, Spectral Interferometric
Polarised CARS (SIP-CARS), that is significantly simpler to implement
than previous experimental approaches. It does not have stringent
requirements on the lasers used, and is suitable for narrowband, multiline
and broadband systems. Transform limited pulses are not required and
PCF generated supercontinua can therefore be used to generate broad,
NRB-free, vibrational spectra. Furthermore, the NRB is removed without
requiring an independent measurement of its spectral variation. The
technique is similar to Dual Quadrature Spectral Interferometry (DQSI)\citep{Lepetit1995,Lim2005,Evans2004,Wipfler2012},
except that the fields have different frequencies and hence only interfere
via the non-linear response. As a result, the two quadratures measured
in SIP-CARS (i.e. the real and imaginary components of the non-linear
response) contain different linear combinations of the tensor elements
of $\chi$, due to polarisation mixing. The third-order response is
solved exactly, without assumptions on the relative strength of resonant
and non-resonant components\citep{Lim2005,Langbein2009}.

To illustrate the method, consider CARS with a (right-hand) circularly
polarised Stokes, and pump and probe linearly polarised along the
\emph{x-}axis, as in Figure \ref{fig:1}(b). For convenience, we express
the Stokes electric field in terms of the equivalent linear polarisation,
$E_{S}$, used to generate it via a $\nicefrac{\lambda}{4}$ plate;
so, the field along \emph{x}- and \emph{y}-axes is $E_{S_{x}}=E_{S}/\sqrt{2}$
and $E_{S_{y}}=iE_{S}/\sqrt{2}$ (along any two orthogonal directions
the Stokes electric field will differ in phase by $\nicefrac{\pi}{2}$).
The susceptibility can be separated into diagonal (e.g. $\chi_{_{iiii}}$)
and off-diagonal (e.g. $\chi_{_{ijij}})$ elements; the off-diagonal
terms mediate the coupling of orthogonally polarised excitations into
the detected polarisation. Considering the polarisation, $P_{+}$,
induced in the medium along the $\hat{\mathbf{n}}_{+}$ axis, at $+\nicefrac{\pi}{4}$
to the pump and probe polarisation, we have
\begin{eqnarray*}
P_{+} & = & \frac{1}{2}\chi_{_{1111}}E_{pr}E_{p}(E_{S}^{+})^{*}+\frac{1}{2}\chi_{_{1221}}E_{pr}E_{p}(E_{S}^{+})^{*}\\
 &  & +\frac{1}{2}\chi_{_{1122}}E_{pr}E_{p}(E_{S}^{-})^{*}+\frac{1}{2}\chi_{_{1212}}E_{pr}E_{p}(E_{S}^{-})^{*}
\end{eqnarray*}
where $E_{p}$ and $E_{pr}$ are the electric fields of the pump and
probe, and $E_{S}^{+}$ and $E_{S}^{-}$ are the components of the
Stokes along $\hat{\mathbf{n}}_{+}$ and $\hat{\mathbf{n}}_{-}$,
respectively (Fig. \ref{fig:1}(b)). As $E_{S}^{+}=iE_{S}^{-}$, the
last two terms lag the first two in phase by $\nicefrac{\pi}{2}$.
Imaginary components of the last two terms therefore interfere with
real components of the first two, with the strength of the interference
determined by the relative strength of the diagonal and off-diagonal
tensor elements. Similarly, for $P_{-}$ the $E_{S}^{+}$ and $E_{S}^{-}$
terms are swapped, and the last two terms lead the first two by $\nicefrac{\pi}{2}$.
Imaginary components are therefore added to the real components along
$\hat{\mathbf{n}}_{+}$, and subtracted from them along $\hat{\mathbf{n}}_{-}$;
spectral interferometric detection is performed by taking the difference
between spectra measured at these polarisations, leaving the purely
imaginary components \citep{Lim2005}.

The induced polarisations are more succinctly expressed within a basis
including the pump and probe polarisations,
\begin{equation}
\begin{array}{c}
P_{x}=\frac{1}{\sqrt{2}}\chi_{_{1111}}E_{pr}E_{p}E_{S}^{*}\\
P_{y}=-\frac{i}{\sqrt{2}}\chi_{_{2112}}E_{pr}E_{p}E_{S}^{*}
\end{array}.\label{eq:Px Py}
\end{equation}

Along the detection axes $\hat{\mathbf{n}}_{+}$ and $\hat{\mathbf{n}}_{-}$
the induced polarisations are $P_{+}=\frac{1}{\sqrt{2}}(P_{x}+P_{y})$
and $P_{-}=\frac{1}{\sqrt{2}}(P_{x}-P_{y})$, and the anti-Stokes
signals are $S_{+}\propto P_{+}P_{+}^{*},\: S_{-}\propto P_{-}P_{-}^{*}$.
The sum and difference of the anti-Stokes intensities are then $\begin{array}{c}
\Sigma S=S_{+}+S_{-}\propto P_{x}P_{x}^{*}+P_{y}P_{y}^{*}\end{array}$ and $\Delta S=S_{+}-S_{-}\propto P_{x}P_{y}^{*}+\left(P_{x}P_{y}^{*}\right)^{*}$,
respectively. The difference signal is therefore given by
\begin{equation}
\triangle S\propto\mathrm{Im}\left\{ \chi_{_{1111}}\chi_{_{2112}}^{*}\right\} I_{pr}I_{p}I_{S}\label{eq:Si-Sj with full chi}
\end{equation}
where $I_{i}=E_{i}E_{i}^{*}$ are the beam intensities. Separating
the susceptibilities into resonant and nonresonant components $\chi_{_{ijkl}}=\chi_{_{ijkl}}^{NR}+\chi_{_{ijkl}}^{R}$,
assuming an isotropic medium $(\chi_{_{2112}}=$$\chi_{_{1221}},\:\chi_{_{1111}}=\chi_{_{1212}}+\chi_{_{1221}}+\chi_{_{1122}})$,
and noting that the non-resonant terms possess Kleinman symmetry\citep{Kleinman1962}
$(\chi_{_{1111}}^{NR}=\chi_{_{NR}},\:\chi_{_{1212}}^{NR}=\chi_{_{1122}}^{NR}=\chi_{_{1221}}^{NR}=\chi_{_{NR}}/3)$
this becomes
\begin{equation}
\begin{array}{r}
\triangle S\underset{}{\propto}\chi_{_{NR}}\mathrm{Im}\left\{ \chi_{_{1111}}^{R}-3\chi_{_{1221}}^{R}\right\} I_{pr}I_{p}I_{S}\\
\underset{}{\propto}(1-3\rho)\chi_{_{NR}}\mathrm{Im}\left\{ \chi_{_{1111}}^{R}\right\} I_{pr}I_{p}I_{S}
\end{array}\label{eq: final diff expression}
\end{equation}
where $\rho=\chi_{_{1221}}^{R}/\chi_{_{1111}}^{R}$ is the CARS depolarisation
ratio of the resonance \citep{Gachet2006}. The difference spectrum
is therefore linear in the imaginary component of $\chi$, is amplified
by $\chi_{_{NR}}$, and contains no real, dispersive terms or non-resonant
contributions. Due to the linear response, well established linear
multivariate analyses such as principal component analysis can be
applied. Mode symmetry information is mixed into the amplitude response
through the depolarisation ratio, $\rho$ ($0<\rho<\nicefrac{3}{4}$),
and the spectrum can be compared directly to polarised spontaneous
Raman spectra. 

Because the interference is effectively between $\nicefrac{\pi}{2}$
phase shifted copies of the same fields there are no extra requirements
on the coherence of the excitation pulses. Moreover, if spectra are
measured simultaneously, incoherent backgrounds (such as two-photon
fluorescence) and any variation of the real components of the CARS
signal are common-mode in $S_{+}$ and $S_{-}$ and are automatically
subtracted out (though they will still contribute shot noise to the
difference spectrum). This is essentially a balanced homodyne detection
scheme, except that in this case the signal arises in the low noise
difference channel rather than the sum channel.

At the focus of a high NA lens it is easier to control the circular
polarisation of a narrowband beam rather than a broadband one, and,
commonly, the pump and probe fields are supplied by the same beam,
so $E_{pr}=E_{p}$. To address this we can generalise (\ref{eq: final diff expression})
to the case of arbitrary ellipticity of both Stokes and pump beams
\footnote{See Supplemental Material at {[}URL will be inserted by publisher{]}
for derivation of the SIP-CARS amplitude for arbitrary ellipticities.%
} 
\begin{equation}
\triangle S\propto C(\theta,\phi)I_{p}^{2}I_{S}(1-3\rho)\chi_{_{NR}}\mathrm{Im}\{\chi_{_{1111}}^{R}\}\label{eq: general diff eqn}
\end{equation}

where
\begin{equation}
C(\theta,\phi)=\frac{1}{2}[\sin(4\phi)+\sin(2\theta)+\sin(2\theta)\cos(4\phi)].\label{eq: C(theta,phi)}
\end{equation}
$\theta$ and $\phi$ characterise the ellipticity of the Stokes and
pump/probe fields, respectively, and are defined as the angle between
the fast axis (set parallel to the \emph{x}-axis) of a quarter waveplate
and an input linear polarisation. For the experimentally practical
situation where the broadband Stokes is constrained to be linear (ie
$C(0,\phi)$) the SIP-CARS signal, $\triangle S$, is maximised for
an elliptical pump with $\phi=\nicefrac{\pi}{8},\nicefrac{3\pi}{8}$.
Significantly, $\triangle S$ retains the same spectral form regardless
of the ellipticity of the excitation beams\emph{.} In general, the
NRB will be removed as long as the polarisation ellipses are symmetric
with respect to the measurement axes, while the ellipticity determines
the amplitude of $\triangle S$. This decoupling of NRB removal and
signal amplitude simplifies the alignment under tight focussing conditions.
Requirements on polarisation purity are somewhat relaxed and the polarisation
can be set \emph{in situ} at the focus by iteratively minimising the
NRB and maximising the difference signal at resonance.

\begin{figure}
\begin{centering}
\includegraphics{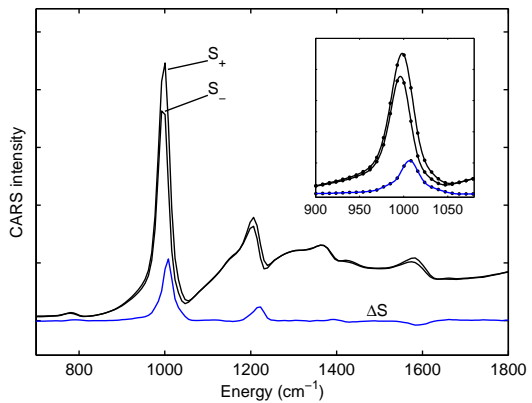}
\par\end{centering}

\caption{(colour online) Interferometric correction of the NRB in toluene.
$S_{+}$ and $S_{-}$ are the raw CARS spectra measured along the
$\hat{\mathbf{n}}_{+}$ and $\hat{\mathbf{n}}_{-}$ directions of
Fig. \ref{fig:1}(b). The difference spectrum $\Delta S=S_{+}-S_{-}$
contains non-dispersive lineshapes at the correct Raman shifts (c.f.
Fig. \ref{fig: cyclo comp w spont R}). Inset: close-up of the region
around 1000 cm$^{-1}$ showing the $\Delta S$ peak shifted with respect
to the dispersive CARS peaks (lines are a guide to the eye). \label{fig:Toluene NRB removal}}
\end{figure}

Experiments with an elliptical pump and broadband linear Stokes were
performed with the apparatus detailed in Figure \ref{fig:1}(c). Dispersion
in the long internal PCF of the source limited temporal overlap between
pump/probe and Stokes, such that the effective Stokes range was 750-2300
cm$^{-1}$ (though strong resonances were detectable to 3000 cm$^{-1}$
\footnote{See Supplemental Material at {[}URL will be inserted by publisher{]}
for comparison of the full SIP-CARS and spontaneous Raman spectra.%
}). The pump/probe width was 30 cm$^{-1}$. This system served as a
rigorous demonstration of the robustness of the technique, as the
PCF generated Stokes beam has relatively low spectral coherence, and
all the wavelengths used were outside the design range for the input
objective lens which significantly affected SNR through poor focussing
performance. Single-shot broadband interferometric NRB removal for
this system would not be possible by any other optical technique. 

Correction of Raman lineshapes and removal of the NRB in SIP-CARS
is shown for toluene in Figure \ref{fig:Toluene NRB removal}. The
two CARS spectra, $S_{+}$ and $S_{-}$, measured at $\pm45^{\mathrm{o}}$
from the Stokes polarisation exhibit the asymmetric dispersive lineshapes
and spectrally varying NRB (the variation reflecting the spectrum
of the Stokes beam) which is characterstic of CARS measurements. The
difference spectrum, $\Delta S$, shows no NRB and Raman peaks are
symmetric and occur at the correct vibrational energy. The inset shows
this in greater detail for the ring breathing mode at 1004 cm$^{-1}$
\citep{Wilmshurst1957}.

Direct comparison to spontaneous Raman spectra can be made by equating
the depolarisation ratio in Equation \ref{eq: final diff expression}
with the spontaneous Raman depolarisation ratio\citep{Gachet2006,YURATICH1977}
$\rho=R_{\bot}/R_{\Vert}$, where $R_{\Vert}$ and $R_{\bot}$ are
spontaneous Raman spectra with incident and scattered polarisation
mutually parallel and perpendicular, respectively. Then $R_{\Vert}\propto\mathrm{Im}\{\chi_{1111}\}$
and $R_{\bot}\propto\mathrm{Im}\{\chi_{1221}\}$, and from Equation
\ref{eq: final diff expression},
\begin{equation}
\frac{\Delta S}{I_{S}}\propto R_{\Vert}-3R_{\bot}.\label{eq:normed diff spectrum}
\end{equation}

\begin{figure}
\begin{centering}
\includegraphics{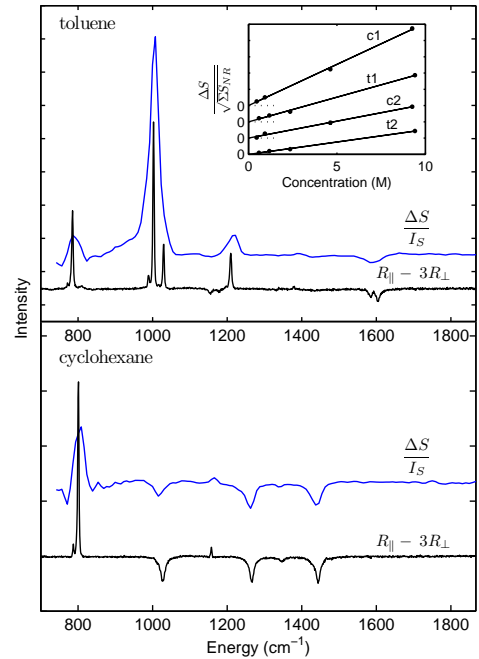}
\par\end{centering}

\caption{\label{fig: cyclo comp w spont R}(colour online) Comparison of normalised
SIP-CARS to the spontaneous Raman spectrum formed by $R_{\parallel}-3R_{\bot}$
($R_{\parallel}$ parallel polarised, $R_{\perp}$ perpendicular polarised).
Curves offset for clarity. Top panel: cyclohexane. Bottom panel: toluene.
The SIP-CARS and spontaneous Raman spectra agree closely, to within
the resolution of the SIP-CARS measurement. Inset: Concentration dependence
of various resonances of toluene (in methanol: 100\% to 6.25\% toluene)
and cyclohexane (in toluene: 100\% to 5\% cyclohexane) in the presence
of stongly varying NRB (see text). $\Sigma S_{NR}$ is $\Sigma S$
integrated over the range 1800-1980 cm$^{-1}$. Error bars are within
marker size. Curves have been offset and scaled for clarity (toluene:
t1=1004 cm$^{-1}$, t2=1208 cm$^{-1}$; cyclohexane: c1=1267 cm$^{-1}$,
c2: 1444 cm$^{-1}$).}
\end{figure}

Note that the normalisation of the difference spectrum by $I{}_{S}$
corrects the peak amplitudes for variation in the Stokes spectrum
to allow comparison with the spontaneous Raman spectrum; it is not
necessary for removing the NRB or for quantitative measurements. For
a non-resonant sample, $\chi_{_{1111}}=\chi_{NR}$ and $\chi_{_{1221}}=\chi_{NR}/3$,
so (\ref{eq:Px Py}) reduces to $P_{x}=\chi_{NR}E_{p}^{2}E_{S}^{*}/\sqrt{2},\; P_{y}=-i\chi_{NR}E_{p}^{2}E_{S}^{*}/3\sqrt{2}$
(where we have set $E_{pr}=E_{p}$). The sum spectrum, $\Sigma S$,
is then 
\begin{equation}
\Sigma S\propto P_{x}P_{x}^{*}+P_{y}P_{y}^{*}=\frac{5}{9}I_{p}^{2}I_{S}\chi_{NR}^{2}.\label{eq:non-res sum}
\end{equation}
The spectral form of $I_{S}$ can therefore be determined from that
of $\Sigma S$ obtained from a non-resonant medium; we have employed
a glass coverslip for this purpose as the glass resonant response
is slowly varying and weakens towards higher wavenumbers. We find
complete agreement between SIP-CARS and spontaneous Raman in terms
of both spectral position and relative peak heights (shown for cyclohexane
and toluene in Figure \ref{fig: cyclo comp w spont R}). As expected
from (\ref{eq: general diff eqn}), the amplitude of each Raman line
is scaled by the depolarisation ratio, with peaks going negative for
resonances with $\rho>\nicefrac{1}{3}$, providing a powerful approach
for the differentiation of otherwise similar spectra on the basis
of mode symmetry. Modes for which $\rho\simeq\nicefrac{1}{3}$ do
not appear, such as the CH$_{3}$ 'umbrella' deformation mode of toluene
at 1379 cm$^{-1}$.

As the non-resonant response, $\chi_{NR}$, amplifies the resonant
CARS signal, quantitative measurements in heterogeneous media require
account to be taken of any variation of $\chi_{NR}$; for example,
as density changes across a sample. SIP-CARS is self-calibrating,
in that $\chi_{NR}$ can be monitored in the same measurement if a
spectrum contains a non-resonant region (such as the 'quiet' region
exhibited by biological samples). From (\ref{eq:non-res sum}), measuring
$\Sigma S$ at a reference frequency away from resonances gives a
quantity $\Sigma S_{NR}\propto\chi_{NR}^{2}$. The SIP-CARS difference
spectrum can therefore be normalised by $\sqrt{\Sigma S_{NR}}$ to
give a signal which is linear in concentration and independent of
the strength of the non-resonant response.

We find that the NRB signal from toluene is approximately twice as
strong as for cyclohexane and much stronger than for methanol. Therefore,
to demonstrate linear concentration dependence in a situation with
varying NRB, a range of mixtures of cyclohexane in toluene and toluene
in methanol were investigated. For cyclohexane in toluene the NRB
therefore increased with decreasing cyclohexane concentration, while
for toluene in methanol the NRB had the opposite trend. $\Sigma S_{NR}$
was measured by integrating the sum spectrum over the range 1800-1980
cm$^{-1}$, and used to normalise the SIP-CARS measurement, which
displayed a linear signal dependence on concentration (inset of Figure
\ref{fig: cyclo comp w spont R}).

In summary, by exploiting the third-order polarisation response, SIP-CARS
allows acquisition of CRS spectra free of NRB, with complete agreement
to spontaneous Raman measurements. Spectra are amplified by the non-resonant
respone and are quantitative, with a linear concentration dependence.
The method uses only passive polarisation optics, has low stability
requirements, and is suitable for any laser system capable of generating
CARS, permitting single-shot interferometric NRB removal with broad
PCF generated supercontinua. 
\begin{acknowledgments}
This work was supported by the Biotechnology and Biological Sciences
Research Council (BB/F016344) and the European Metrology Research
Programme (NEW02-REG2).
\end{acknowledgments}
\bibliographystyle{apsrev}
\bibliography{CARSbib}

\end{document}